\documentclass{cernyrep}
\usepackage{float}
\usepackage[T1]{fontenc}
\usepackage[bookmarks, colorlinks=true, linktoc=page, pdftex, linkcolor=black, citecolor=black, urlcolor=blue]{hyperref}

\usepackage{fancyhdr}
\usepackage{empheq}
\fancyhfoffset{4 mm}
\fancypagestyle{ARTTITLE}{% 
\fancyhf{} % clear all header and footer fields
\lhead{\hfill CERN Accelerator School Proceedings ---
{\it RF for Accelerators} ---  Berlin,  Germany, 2023\hfill}
\lfoot{\hspace{3mm} Available online at \url{https://cas.web.cern.ch/previous-schools}}
\rfoot{\thepage\hspace*{3mm}}
 
}

\usepackage{varwidth}
\usepackage{xcolor}
\usepackage{graphicx} % Required for inserting images
\usepackage[
    backend=biber,
    style=numeric,
    sorting=none,
    giveninits=true
  ]{biblatex} %Imports biblatex package
\usepackage{xurl}
\usepackage{hyperref}

\DeclareFieldFormat{url}{\newline\url{#1}}

\usepackage{caption}
\usepackage{graphicx}
\usepackage{subcaption}

\title{HOM Mitigation}
\author{Nicoleta Baboi, DESY, Hamburg, Germany}
\date{\today}

\begin{document}
%\maketitle

\begin{abstract}
    Higher Order Modes (HOMs) in accelerating structures are resonant electromagnetic fields excited and left behind by charged particle bunches. 
They can severely degrade the beam quality in accelerators, such as the multi-bunch energy spread and transverse emittance, and therefore one must carefully study and mitigate them.
In this lecture, various methods of HOM mitigation are reviewed. 
On one hand, one tries to avoid their existence, for example by careful design of the accelerating structures and damping systems. 
On the other hand, one can reduce the effect of the excited HOM fields, for example by means of fast feedbacks or beam alignment. 
Additionally, a few ways to make HOM signals useful are referred to, such as diagnostics of the beam or of accelerating structures. 
\end{abstract}

\maketitle
\thispagestyle{ARTTITLE}

%========================================
%\input{sections/1_Introduction}
\section{Introduction}\label{sec:intro}

Before reading these lecture notes, take a moment to recall what you already know about Higher Order Modes (HOMs) in accelerators.
You may want to restrict yourself to what you have learned at this school, such as in the lecture on ``Impedances and Wakefields" by Andrea Mostacci \cite{mostacci_impedances_2023}.
At the end you may want to come back to this question and assess what was new to you, what is relevant for your work, and what you might want to explore in greater detail. 

This lecture is by no means exhaustive, nor is the list of references, which serves merely as a~starting point for further research on the topic. 
Its goal is to provide an overview of the subject and suggest some starting points for those who wish to explore it further.

\subsection{Wakefields}
The term Higher Order Mode is strongly intertwined with the term wakefield.
Let us first look at the~latter. 
The Merriam-Webster dictionary defines a \textit{wake}, among other meanings, as \cite{mw:wake}:
\begin{enumerate}
    \item {the track left by a moving body (such as a ship) in a fluid (such as water); broadly: a track or path left.}
    \item {aftermath (the period immediately following a usually ruinous event).}
\end{enumerate}

\begin{figure}[ht]
    \centering
       \begin{subfigure}[b]{0.58\textwidth}
       \centering
       \includegraphics[width=\linewidth]{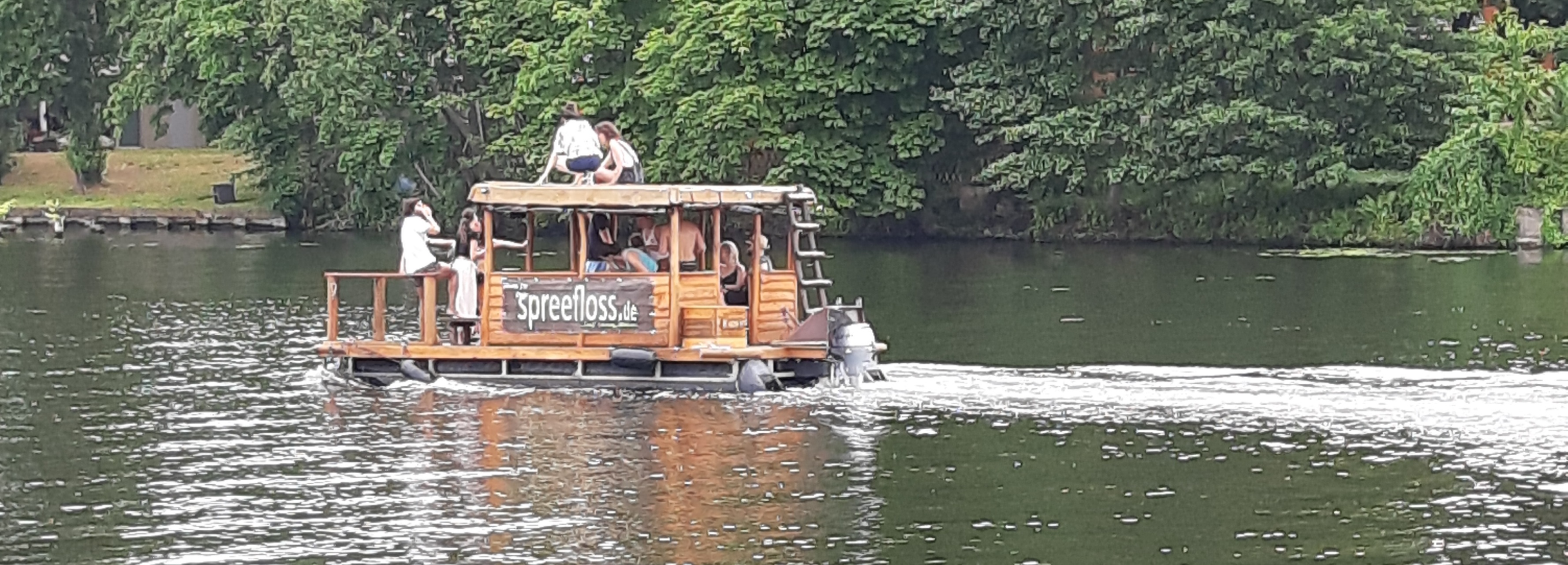}
       \caption{}
       \label{fig:wake_boat}
    \end{subfigure} 
    \hfill
    \begin{subfigure}[b]{0.37\textwidth}      
       \centering
       \includegraphics[width=\linewidth]{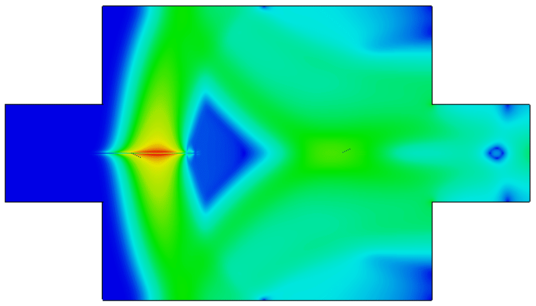}
       \caption{}
       \label{fig:wake}
    \end{subfigure}
\caption{The wake of a ship on a river (a); Wakefield excited by a charged particle in a resonant cavity (b).}
\label{fig:wakefield}
\end{figure}

An example of the first definition was experienced at this school's boat trip on the Spree River, as shown in Fig.~\ref{fig:wake_boat}.
In accelerator physics, the ``moving body" is not a ship but a charged particle or a~bunch of particles, and the  ``track'' consists of electromagnetic fields. 
Thus we refer to it as \textit{wakefield}.

What about the second definition? 
Are wakefields ``following a usually ruinous event"? 
Not necessarily, but they can themselves be the destructive event, leading to an ``aftermath". 

Figure~\ref{fig:wake} shows an example of a wakefield in a simple resonant cavity\footnote{Courtesy of Sosoho-Abasi Udongwo}. 
When a charged particle traverses a perfectly conducting beam pipe, the accompanying electromagnetic fields propagate alongside it. 
One can conceptualize an image charge traveling parallel to the particle in the beam pipe, such that  the field lines connect the two. 
If the beam pipe has an opening, as introduced by a resonant cavity like the one in the figure, the image charge must follow a longer path than the particle.
This results in a~deformation of the field lines.
After the particle exits the cavity, part of its electromagnetic fields remain trapped inside, resonating  for a while \cite{mostacci_impedances_2023}. 

Apart from geometrical variations in the beam pipe, as in this example, wakefields can also arise due to the finite resistance of the beam pipe walls, or even the roughness of its surface.
%*** moved from below
Only geometrical fields are relevant for this lecture.
Also,  we will focus primarily on RF accelerating structures, though the~formalism applies equally to other accelerator components.
At the same time, unless otherwise stated, terms such as ``accelerating cavity'' or ``structure'' also encompass other types of active RF cavities, such as crabbing or deflecting cavities.
The primary distinction lies in the resonant mode employed: monopole modes, with longitudinal electric field are used for acceleration, 
whereas transverse modes (e.g., dipole) are utilized for bunch crabbing or deflection.
It is also important to note that in general we restrict to \textbf{circularly symmetrical structures} used in most cases.
% ------------------------

Wakefields are generally classified as either longitudinal or  transverse.
One also differentiates between \textbf{short-range wakefields}, on the order of the bunch length, and \textbf{long-range wakefields}, along a~bunch train. 
Figure~\ref{fig:wake} showed an example of a short-range wakefield.

\begin{figure}[h]
    \centering
    \includegraphics[width=0.5\linewidth]{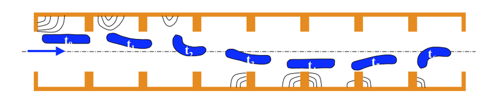}
    \caption{Example of single bunch beam breakup (``banana" effect) in an accelerating structure  \cite{barletta_uspas_2010}.}
    \label{fig:single-wake}
\end{figure}

Figure~\ref{fig:single-wake} depicts a possible single bunch effect in an accelerating structure \cite{barletta_uspas_2010}. 
Consider a bunch entering the structure with an initial transverse offset.
The head of the bunch generates wakefields, which contain transverse components. 
These wakefields act on the tail of the bunch,  imparting a transverse kick. 
As the bunch advances through the structure, the head continues to excite fields that affect the tail, potentially amplifying the effect.
Over time, the bunch develops a characteristic ``banana'' shape, leading to an increase in transverse emittance and, in severe cases, partial beam loss --- a phenomenon known as single-bunch beam breakup.
In the longitudinal plane, short-range wakefields may induce an increase in the bunch energy spread. 
%Short-range wakefields are typically analyzed in time domain.

This lecture focuses on long-range wakefields.
Examples of their effects will be given later in this section.

\subsection{Higher Order Modes}

While short-range wakefields are typically analyzed in the time domain, long-range ones are most often treated in the frequency domain. 

%*** moved above:
%Throughout this lecture, we will focus primarily on RF accelerating structures, though the formalism applies equally to other accelerator components.
%Also, unless otherwise stated, terms such as ``accelerating cavity'' or ``structure'' also encompass other types of active RF cavities, such as crabbing or deflecting cavities.
%The primary distinction lies in the resonant mode employed: monopole modes, with longitudinal electric field are used for acceleration, 
%whereas transverse modes (e.g., dipole) are utilized for bunch crabbing or deflection.
%It is also important to note that in general we restrict to \textbf{circularly symmetrical structures} used in most cases.
%Finally, this discussion generally assumes circularly symmetric structures, which cover most practical cases.
% ------------------------

The \textbf{longitudinal wake potential} (hereafter referred to simply as the wakefield)  can be written as a sum over all eigenmodes of the accelerating cavity \cite{Wanzenberg:415327}:

\begin{equation} \label{eq:Wake_long}
W_{||}(s) = \sum_n 2 k_n cos \left( \omega_n \frac{s}{c} \right) e^{-\frac{\omega_n}{2 Q_n} \frac{s}{c}},
\end{equation}

\noindent 
where $\omega_n$ is the angular frequency, $k_n$ the longitudinal loss factor, \(Q_n\) the quality factor of mode with index \(n\), \(s\) is the distance behind the excitation particle and \(c\) the light speed in vacuum.
One often calls the individual components in this sum Higher Order Modes (HOMs), although strictly speaking the~fundamental mode is also excited by the beam. 
Sometimes also Lower Order and Same Order Modes are present. 
For simplicity, all contributions to the long-range wakefield are denoted as HOMs in this lecture.
Note that in principle also short-range wakefields can be treated as a sum of eigenmodes, however one would need to go to quite high frequencies, which is difficult.

Each mode is characterised by three main parameters: frequency \(f_n = \omega_n / 2 \pi\), loss factor \(k_n\) and quality factor \(Q_n\).
The longitudinal loss factor is given by:

\begin{equation} \label{eq:loss_factor}
k_n = \frac{|V_n(r,\omega_n)|^2}{4 U_n},
\end{equation}

\noindent 
where \(V_n\) is the voltage induced by mode \(n\) on a trailing particle behind the exciting particle, integrated along the whole structure, and \(U_n\) the stored energy. Its units are \(V/pC\). Often, the so-called (longitudinal)  \(R/Q\) factor is used instead of the loss factor. The two quantities are related by:

\begin{equation} \label{eq:RoverQ}
k_n = \frac{\omega_n}{4} \left( \frac{R}{Q} \right)_n.
\end{equation}

\noindent 
Please note that this ``linac'' definition of \(R/Q\) differs by a factor 2 from the ``RF'' definition.

The loss factor (and correspondingly the \(R/Q\)) quantifies the strength of the interaction between an~accelerating structure (or other components) and the beam. 
Whereas a high value is desirable for the~accelerating mode, the opposite holds for HOMs.
A notable feature of the loss factor is its independence from the material the cavity is made of, depending solely on its geometry!

The last term in Eq.~(\ref{eq:Wake_long}) gives the damping of the mode due to energy dissipation, quantified by the~quality factor \(Q_n\):

\begin{equation} \label{eq:Q_0}
Q_n = \omega_n \frac{stored\ energy}{power\ loss}.
\end{equation}

\noindent 
When considering only the material losses of the cavity, one obtains the intrinsic quality factor, \(Q_0\). 
Additional energy losses can occur through the beam pipe or dedicated ports, giving the external quality factor, \(Q_{ext}\).
The total quality factor of a mode is determined by
\begin{equation} \label{eq:Q_total}
\frac{1}{Q_n} = \frac{1}{Q_0} + \frac{1}{Q_{ext}}.
\end{equation}

\begin{figure}[htbp]
  \centering
  \begin{subfigure}{0.48\textwidth}
    \centering
    \includegraphics[height=5cm, keepaspectratio]{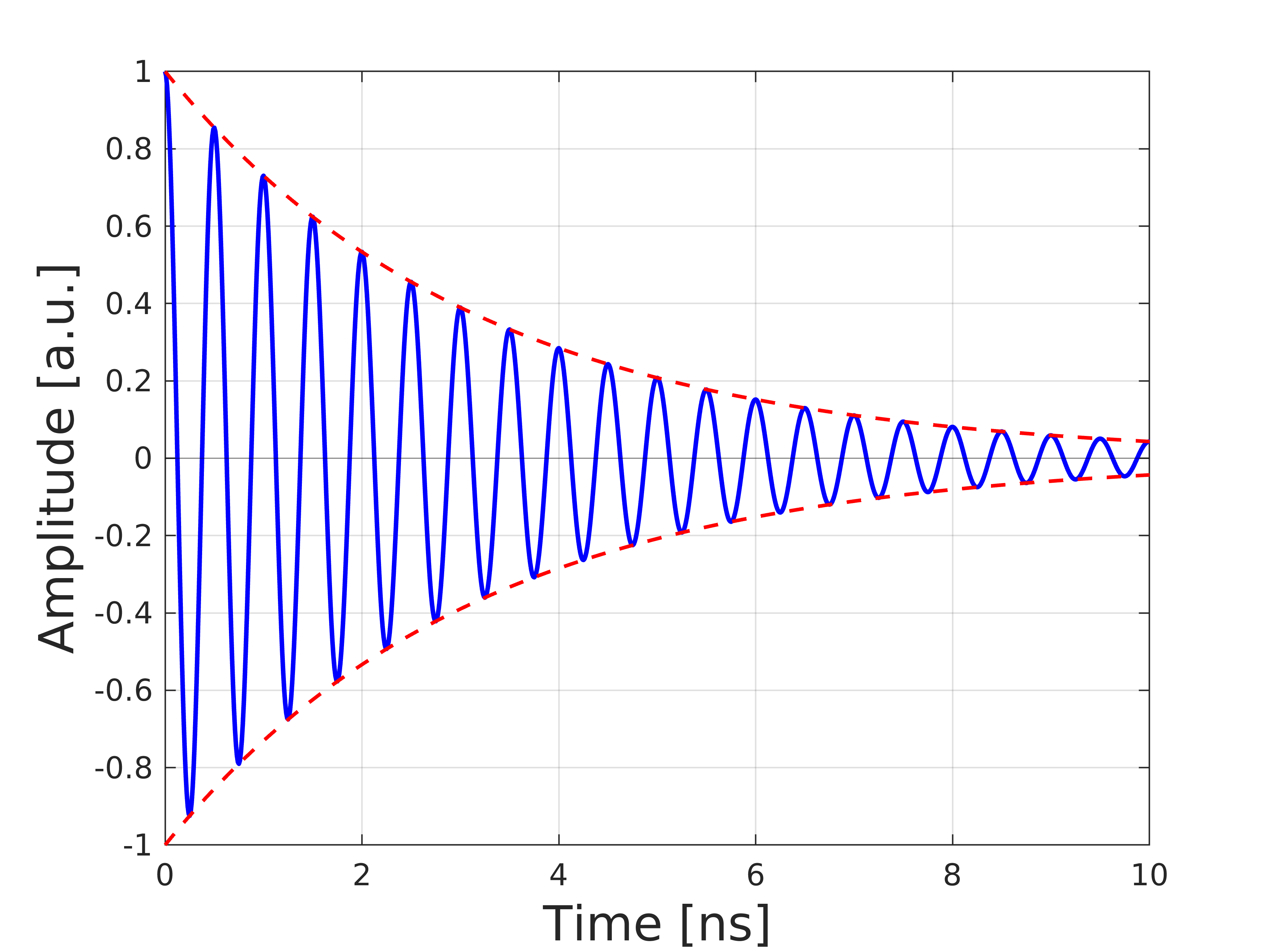} % Fixed height, width auto-adjusted
    \caption{}
    \label{fig:damped_time}
  \end{subfigure}
  \hfill
  \begin{subfigure}{0.48\textwidth}
    \centering
    \includegraphics[height=5cm, keepaspectratio]{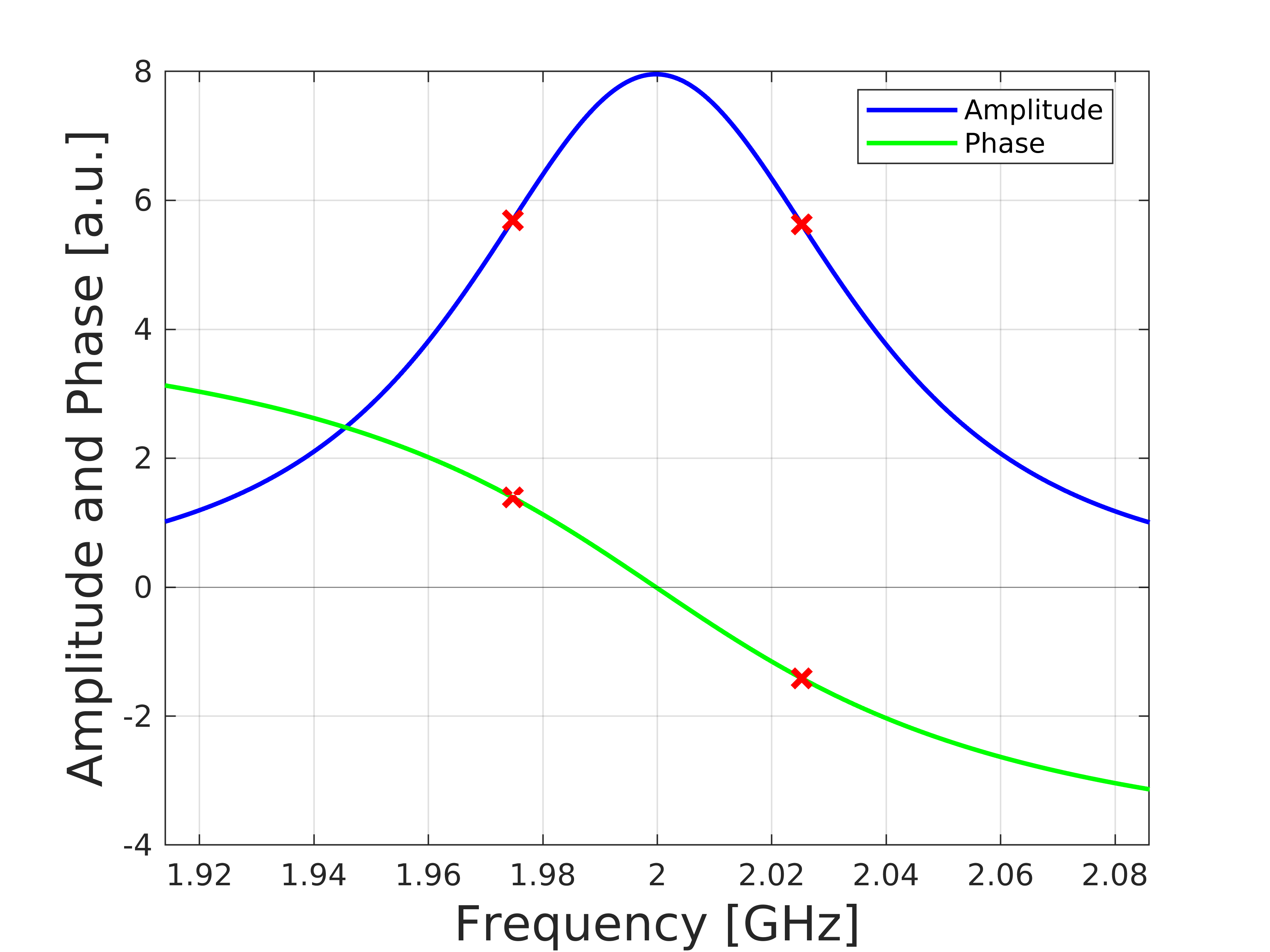} % Same height for both figures
    \caption{}
    \label{fig:damped_freq}
  \end{subfigure}
  \caption{Example of a damped oscillation (a) and its amplitude and phase spectrum (b).}
  \label{fig:damped_resonance}
\end{figure}

Figure~\ref{fig:damped_resonance} qualitatively shows a damped oscillation in time domain (\subref{fig:damped_time}) alongside its amplitude and phase spectrum (\subref{fig:damped_freq}) for a mode with a frequency of 2~GHz.
One observes the exponential damping of the~oscillation, which in frequency domain translates in a certain width of the resonant peak. 
Both the~quality factor and the damping time \( \tau_n \) can be easily determined from the measurement of such a~resonance:

\begin{equation}\label{eq:Q_tau} 
Q_n = \frac{\omega_n}{\Delta \omega_n}; \quad \tau_n = \frac{2 Q_n}{\omega_n},
\end{equation}

\noindent where \( \Delta \omega_n \) is the bandwidth of the resonance.

%*** moved above
%\noindent 
%One observes the exponential damping of the oscillation, which in frequency domain translates in a certain width of the resonant peak. 
%Both the quality factor and the damping time \( \tau_n \) can be easily determined from the measurement of such a resonance:

%\begin{equation}\label{eq:Q_tau} 
%Q_n = \frac{\omega_n}{\Delta \omega_n}; \quad \tau_n = \frac{2 Q_n}{\omega_n},
%\end{equation}
%
%\noindent where \( \Delta \omega_n \) is the bandwidth of the resonance.
% ------------------------

\begin{figure}[h]
  \centering
  \begin{subfigure}{0.2\textwidth}
     \includegraphics[width=\textwidth,height=\textwidth]{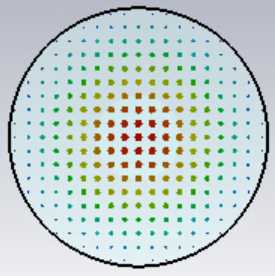} 
     \caption{}
     \label{fig:tm010}
  \end{subfigure}
  \begin{subfigure}{0.2\textwidth}
     \includegraphics[width=\textwidth,height=\textwidth]{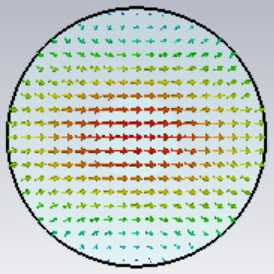}
     \caption{}
     \label{fig:te111_pol1}
  \end{subfigure}
  \begin{subfigure}{0.2\textwidth}
     \includegraphics[width=\textwidth,height=\textwidth]{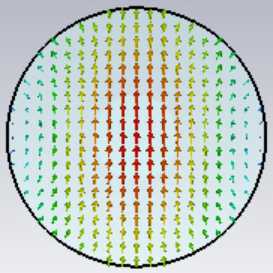}
     \caption{}
     \label{fig:te111_pol2}
  \end{subfigure}
  \caption{Electric field geometry in a simple circularly symmetric resonant cavity for a monopole mode (\subref{fig:tm010}) and the~two polarizations of a dipole mode (\subref{fig:te111_pol1} and \subref{fig:te111_pol2}).}
  \label{fig:monopole_dipole}
\end{figure}

For (quasi-)cylindrically symmetric accelerating structures, the eigenmodes can be categorized into \textbf{monopole}, \textbf{dipole}, \textbf{sextupole} and so on. 
Figure~\ref{fig:monopole_dipole} provides examples of a monopole (\subref{fig:tm010}) and the two polarizations of a dipole mode (\subref{fig:te111_pol1} and \subref{fig:te111_pol2}).
Monopole modes exhibit azimuthal symmetry.
The one shown features a longitudinal electric field with a maximum on the cavity axis --- ideal for particle acceleration.
Dipole modes possess azimuthal nodes at the cavity wall (or one can think of two ``half-nodes'').
Such modes always occur in pairs, which is the case also for other modes with azimuthal asymmetry, e.g.  quadrupoles, sextupoles.
These \textbf{two polarizations} are orthogonal to each other for perfect cylindrically symmetric structures, as in subfigures \subref{fig:te111_pol1} and \subref{fig:te111_pol2}.
In practice, structural asymmetries --- often arising due to manufacturing imperfections, but mainly due to couplers --- affect this orthogonality and split the~resonance frequencies of the two polarizations.

While all mode types contribute to longitudinal beam effects (e.g. change in the energy spread), monopole modes do not induce transverse effects for on-axis beams. 
Transverse effects are in many cases driven by dipole modes.
Similarly to the longitudinal case, the \textbf{dipole transverse wakefield} can be expressed as a sum over HOMs:

\begin{equation}\label{eq:Wake_transv}
W_\perp(s) = \sum_n 2 k'_{n\perp} sin \left( \omega_n \frac{s}{c} \right) e^{-\frac{\omega_n}{2 Q_n} \frac{s}{c}},
\end{equation}

\noindent 
where \( k'_{n\perp} \) denotes the dipole kick factor, defined as:

\begin{equation} \label{eq:k_perp}
k'_{n\perp} = \frac{c k_n}{\omega_n r^2} \quad \text{[V/pC/mm]}.
\end{equation}

\noindent 
As it can be seen, the loss factor is here normalized to the offset squared. 
This normalization is practical because the excited electric field of dipole modes scales linearly with the offset, so that the kick factor is independent of \(r\). 
Equation~(\ref{eq:k_perp}) results from the Panofsky-Wenzel theorem, implying that it is sufficient to calculate the longitudinal loss factor (or R/Q) of a mode. 
Note that dipole modes are not excited by centered beams, so that one needs to calculate the loss factor at an offset, in order to be able to determine the kick factor.
Nonetheless, the excited fields and their effects usually have on-axis components!

While \(f_n\), \(k_n\), \(Q_n\) and, for transverse modes, the polarization tell us a lot about a HOM, further characteristics may be necessary, such as the field distribution, in order to understand its behaviour.
Examples of eigenmodes for some cavities can be found for example in Refs.~\cite{Wanzenberg:415327} and \cite{carter:linac04-thp83}.

In \textbf{multicell cavities} the resonances of each unit couple to each other.
Each mode from a single cell resonator splits into as many modes as cells forming a band of modes.
Coupled RLC circuits are often used to model the behaviour of eigenmodes.
Note that the number of modes may be different from the~number of cells due to the influence of the beam pipes, couplers or nearby components.

\begin{figure}[H]
   \centering
   \begin{subfigure}{0.46\textwidth}
      \centering
      \includegraphics[width=\textwidth]{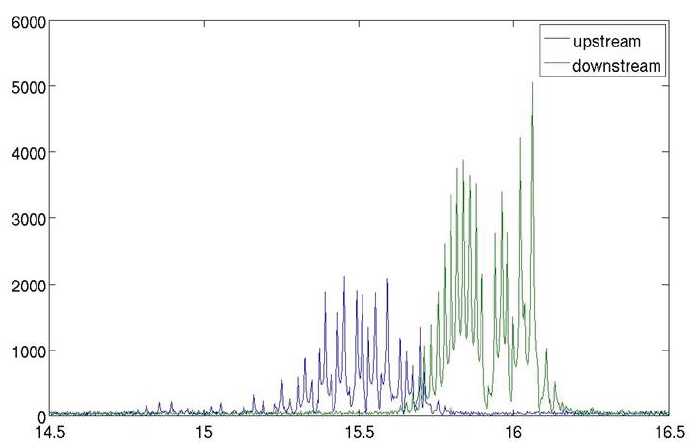}%
      \caption{}
      \label{fig:swissfel-spectrum}
   \end{subfigure}
   \begin{subfigure}{0.46\textwidth}
      \centering
      \includegraphics[width=\textwidth]{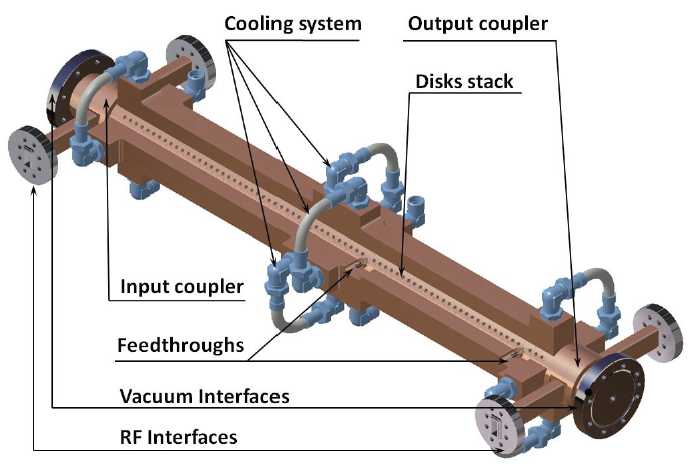}%
      \caption{}
      \label{fig:swissfel}
   \end{subfigure}
   \begin{subfigure}{0.46\textwidth}
      \centering
      \includegraphics[height=5cm]{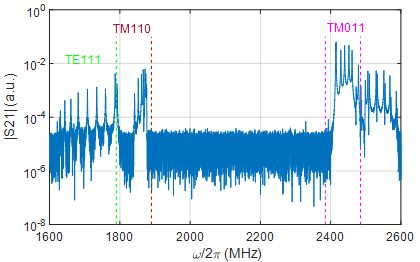}%
      \caption{}
      \label{fig:tesla-cav-spectrum}
   \end{subfigure}
   \begin{subfigure}{0.46\textwidth}
      \centering
      \includegraphics[height=4cm]{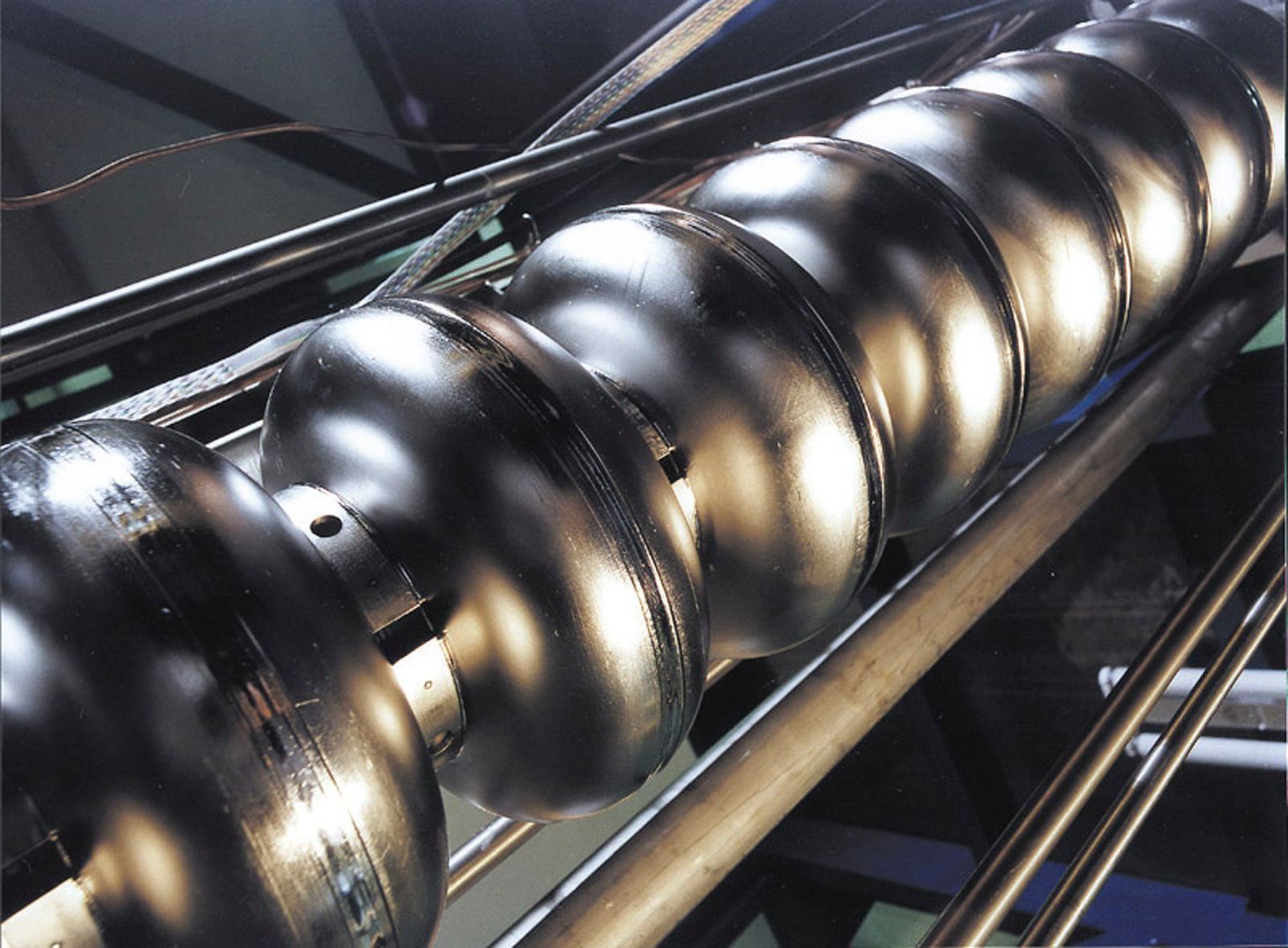}%
      \caption{}
      \label{fig:tesla-cav}
   \end{subfigure}
   \caption{Measured HOM spectrum of the first dipole band (a) of a normal-conducting detuned X-band 72-cell accelerating structure (b)  \cite{dehler_x-band_ibic2013}; Measured HOM spectrum with 2 dipole and 1 monopole band (c) of a SC TESLA cavity (d) \cite{wei_homsc2018}.} 
\end{figure}

Figure~\ref{fig:swissfel-spectrum} shows the measured HOM spectrum of the first dipole band of a normal-conducting (NC) X-band 72-cell accelerating structure (\ref{fig:swissfel}) \cite{dehler_x-band_ibic2013}.
Due to the cell detuning needed to keep the gradient constant, each mode is trapped in part of the~structure.
Each mode can be detected via so-called wakefield monitors positioned upstream and downstream. 
For the superconducting (SC) 9-cell TESLA cavity shown in Fig.~\ref{fig:tesla-cav}, the spectrum can be seen at both couplers installed in the beampipes at the~ends of the~structure.
The transmission coefficient \(S_{21}\) measured across the HOM couplers is presented in Fig.~\ref{fig:tesla-cav-spectrum}  \cite{wei_homsc2018}.
The first two dipole bands (TE111-like and TM110-like), the first higher order monopole band (TM011) and further modes can be seen.

\subsection{Issues due to HOMs}

Each bunch in a bunch train excites HOMs and is affected by the fields excited by the previous bunches.
This can significantly degrade the beam quality.

\begin{figure}[h]
    \centering
    \includegraphics[width=0.65\linewidth]{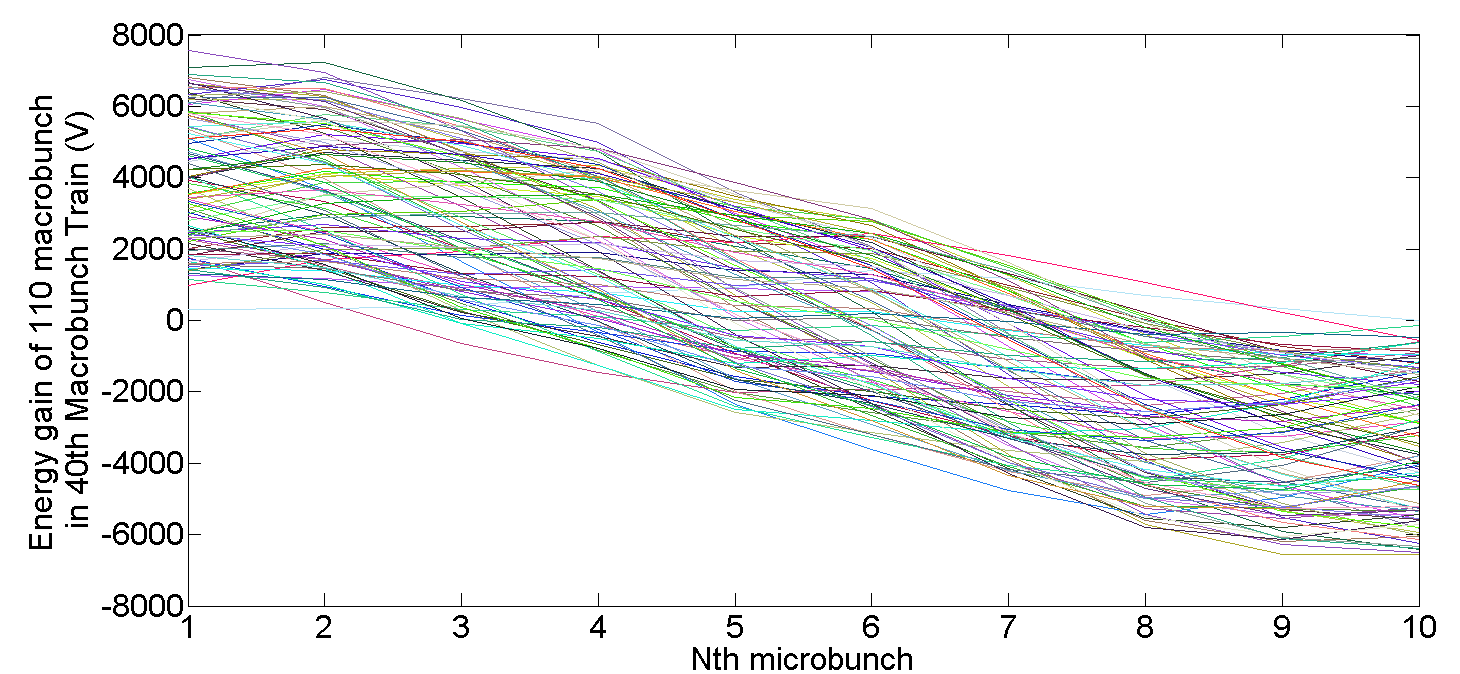}
    \caption{Calculated energy spread in a macropulse estimated for the SC cavity proposed for the eRHIC energy recovery linac \cite{xu_new_2017}.}
    \label{fig:energy-spread}
\end{figure}

In the longitudinal plane the energy spread of the beam is affected.
Figure~\ref{fig:energy-spread} illustrates the simulated energy spread in a macropulse estimated for the SC cavity proposed for the eRHIC energy-recovery linac \cite{xu_new_2017}.

\begin{figure}[h]
   \centering
   \begin{subfigure}{0.5\textwidth}
      \centering
      \includegraphics[height=5cm]{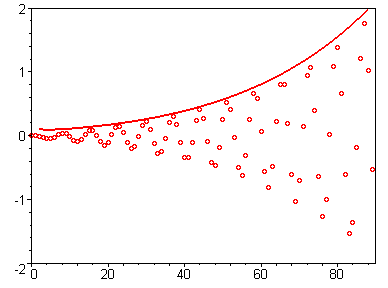}%
      \caption{}
      \label{fig:mbbu}
   \end{subfigure}\hfill
   \begin{subfigure}{0.5\textwidth}
      \centering
      \includegraphics[height=5cm]{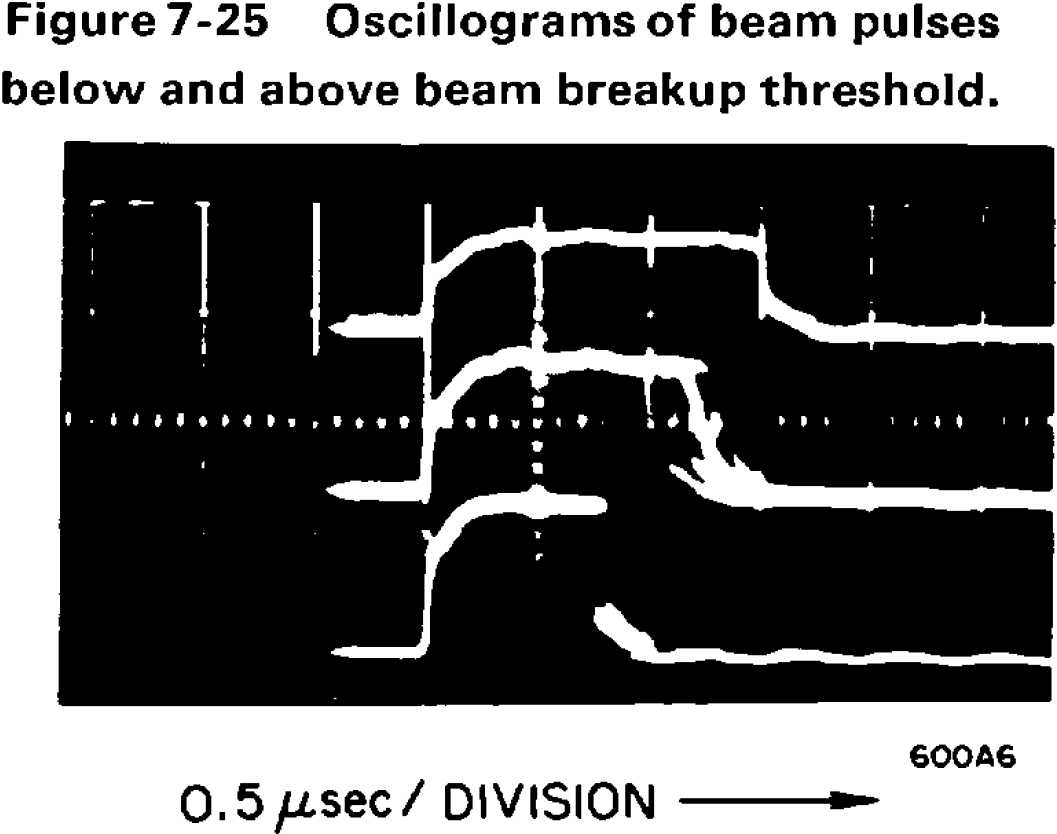}%
      \caption{}
      \label{fig:bbu-slac}
   \end{subfigure}\hfill
   \caption{Simulated exponential increase in the relative bunch offsets at the end of the main linac of the Next Linear Collider study \cite{bohn_theory_2000} (a); Beam loss due to beam breakup observed at the SLAC linac \cite{neal_twomile_1968} (b).}
\end{figure}

In the transverse plane, HOMs can induce a spread in the bunch offsets.
Needless to say that the~kicks on the bunches are stronger at lower energies.
After a while the steady state is usually reached.
However, in the worst-case scenario, the offset may increase along the train, if one or several modes are excited synchronously by each bunch.
An example of this behavior is shown in Fig.~\ref{fig:mbbu}, depicting a simulation for the Next Linear Collider \cite{bohn_theory_2000}.
In extreme cases, multibunch beam breakup can occur, leading to partial beam loss, as historically observed at the Stanford two mile accelerator (Fig.~\ref{fig:bbu-slac}) \cite{neal_twomile_1968}.

Another major concern associated with HOMs is heating caused by the excited fields.
This may lead to RF breakdown in NC structures or quenching in SC cavities.
Moreover, neighboring components may suffer from overheating, as many HOM fields can propagate through the beam pipes.
Diagnostic devices can also be adversely affected if HOM signals interfere with direct beam signals.

When designing a new accelerator facility, it is imperative to mitigate the adverse impacts of HOMs on beam quality.
The following sections outline strategies to avoid exciting resonant modes within accelerating structures (Section~\ref{sec:avoid_homs}), and 
to mitigate their remaining effects (Section~\ref{sec:compensate_homs}). 
Finally, potential beneficial applications of HOMs are briefly addressed (Section~\ref{sec:hom_uses}).

%========================================
%\input{sections/2_Avoid-HOMs}
\section{Methods to Avoid HOM Excitation}\label{sec:avoid_homs}

While HOMs can severely degrade the beam quality in accelerators by increasing energy spread, emittance, or even causing beam breakup, their effects can be reduced in the first place by avoiding their presence in the accelerating structures. On one hand, one can design the structures such that the HOMs are as harmless as possible, or, on another hand, design and operate the accelerator so that the worst modes are not excited.

\subsection{Structure Design}\label{structure-design}
\begin{figure}[h]
    \centering
    \includegraphics[width=0.6\linewidth]{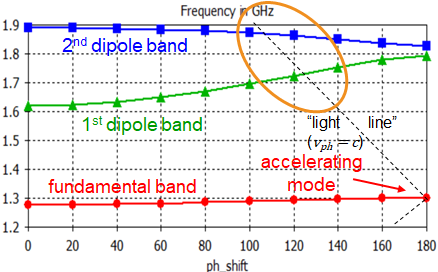}
    \caption{Dispersion diagram of a TESLA cavity (single-cell simulation with periodic boundaries) \cite{Wanzenberg:415327}. Two dipole bands are shown, along with the fundamental band and the ``light line" for ultrarelativistic particles.}
    \label{fig:dispersion_diagram}
\end{figure}

One has to carefully look at the HOMs already when designing an accelerating structure.
By looking at Eqs.~(\ref{eq:Wake_long})  and (\ref{eq:Wake_transv}) one quickly concludes that the long-range wakefields are small when the loss factors of the~summed modes are small.
The target of optimization of the \textbf{cavity geometry} is therefore to minimize the loss factors $k_n$ for HOMs, while ensuring a high value for the accelerating mode. 
Keep in mind that the loss factor depends solely on the geometry. 
In multi-cell structures particular attention has to be paid to the modes which interact strongest to the beam. 
These are the synchronous or quasi-synchronous modes, i.e. the modes with a phase velocity close to the speed of the accelerated particles. 
This can be visualized on the dispersion diagram, giving the frequency of each eigenmode versus its phase advance per cell. 
%*** added some words to previous sentence
Figure~\ref{fig:dispersion_diagram} shows the fundamental band as well as the first two dipole bands on the dispersion diagram for the TESLA cavity used to accelerate relativistic electrons \cite{Wanzenberg:415327}. 
The~so-called ``light line'' helps identify the quasi-synchronous modes, in this case for particles traveling with the speed of light.
Four such dipole modes are marked in the figure, which are the ones to be payed more attention to. 
Indeed the quasi-synchronicity results in higher loss factors (or \(R/Q\), usually used for superconducting cavities). 
Note that of course the accelerating mode is a synchronous mode.

The cavity geometry should be chosen such that the dangerous HOMs are not excited resonantly by each bunch, i.e. the fields generated by each bunch will not add up.
In other words the mode frequency should not coincide with a bunch harmonics.
For long structures with many cells one can opt to detune the~cells along the structure, i.e. vary the cell and iris diameters. 
In this way the excited HOMs will decohere, will not add up; instead, they rather cancel each other out. 
This is typically the case for constant gradient structure in contrast to constant impedance ones \cite{gerigk_cavities_2023}.

Moreover, for accelerators requiring many accelerating cavities one can produce several types of structures with slightly different HOM frequencies, interleaved in each other. 
This will further help the~excited modes to decohere. 
This scenario has been analysed for example for the accelerating structures for CLIC \cite{jones_cas_2010}.
Also the spread in the HOM frequencies caused by fabrication tolerances up to some level are beneficial for reducing their impact on the beam.

Looking again at Eqs.~(\ref{eq:Wake_long})  and (\ref{eq:Wake_transv}), the last term indicates the damping of the wakefield.
The \textbf{material} of the cavity walls is the main factor determining the intrinsic quality factor of the HOMs, \(Q_0\). 
However this is decided by the choice of technology, such that one usually does not have control over it.

What one can control is the loaded (or total) quality factor, which is basically what the beam will see (see Eq.~(\ref{eq:Q_total})). 
By introducing elements which \textbf{extract power from the excited HOMs}, typically couplers, waveguides or absorbers, one damps the mode, i.e. one lowers the Q by adding an external contribution to it.
One has to design this contribution carefully in order not to affect the accelerating mode.
The required level of damping is determined via Beam Break-Up (BBU) models or beam dynamics simulations (see for example Refs.~\cite{shaposhnikova_instabilities_2023}, \cite{lou_beam_2019}, \cite{baboi:epac00-thp3b05}).

\begin{figure}[h]
   \centering
   \begin{subfigure}{0.9\textwidth}
       % \subfloat[]{%
      \centering
      \includegraphics[width=0.8\linewidth]{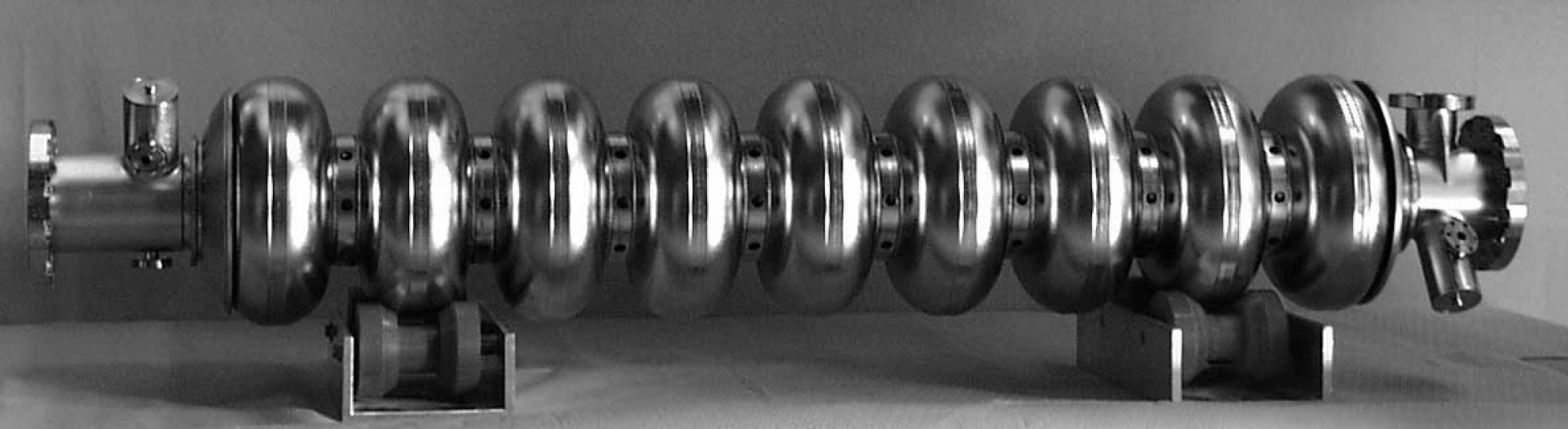}%
      \caption{}
      \label{fig:tesla_cavity}%
       % }\hfill
   \end{subfigure} \hfill
   \begin{subfigure}{0.9\textwidth}      
       % \subfloat[]{%
      \centering
      \includegraphics[width=0.8\linewidth]{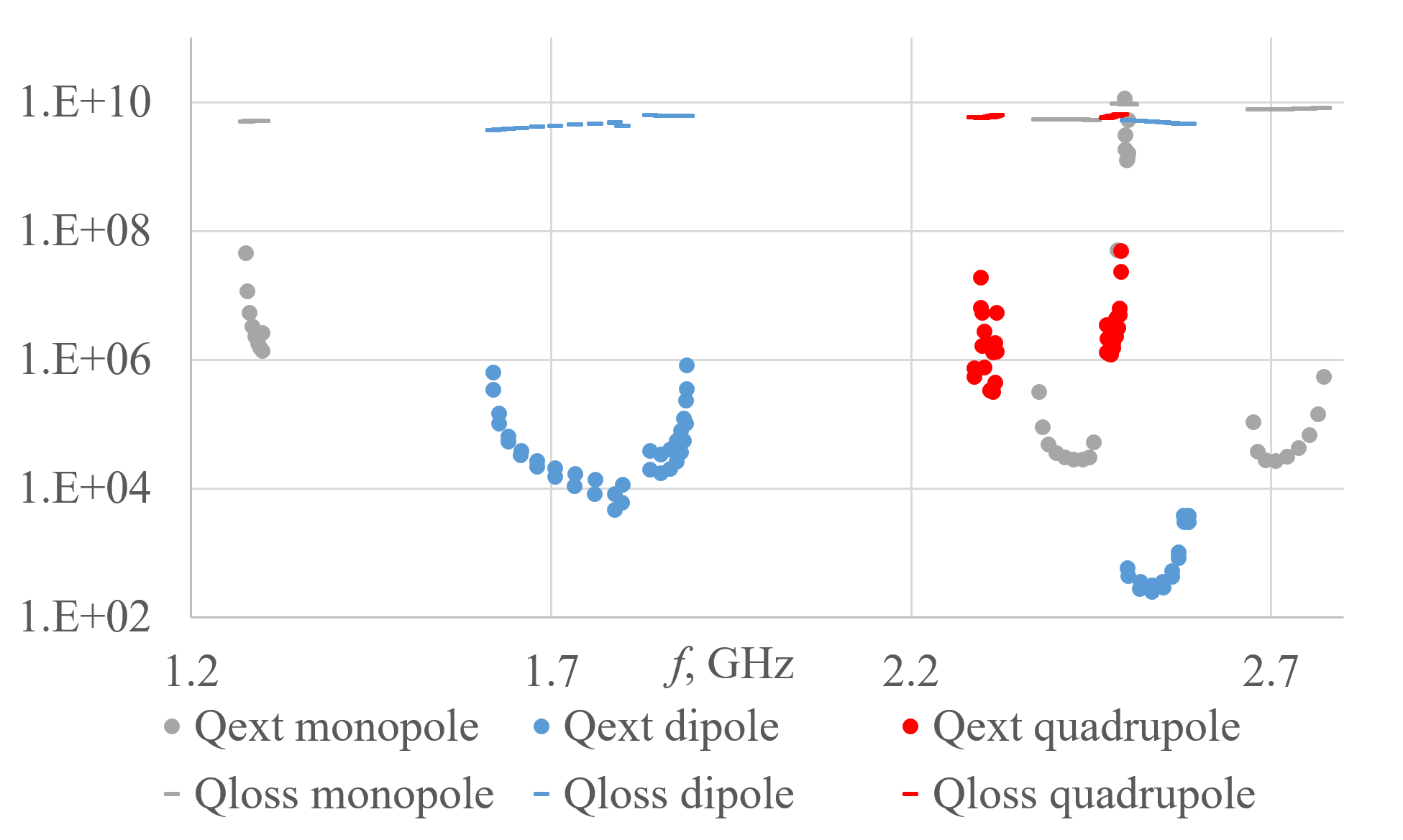}%
      \caption{}
      \label{fig:tesla_cavity_Q}%
       % }\\
   \end{subfigure}
\caption{TESLA cavity with HOM couplers at both ends  (a) and its quality factors $Q_0$ (here named $Qloss$ and $Q_{ext}$ for several monopole, dipole and quadrupole bands (b).}
\end{figure}

The damping effect of \textbf{HOM couplers} placed at either end of a superconducting TESLA cavity (Fig.~\ref{fig:tesla_cavity}) can be seen in Fig.~\ref{fig:tesla_cavity_Q}.
The intrinsic quality factor of each mode \(Q_0\) (denoted \(Q_{loss}\) in the figure) is of the order of 10\textsuperscript{9}, while \(Q_{ext}\), is generally several orders of magnitude lower, leading to \(Q_{total}\) values of the same magnitude.

%*** added a few words in the caption

\begin{figure}[h]
   \begin{subfigure}{0.45\textwidth}
      %  \subfloat[]{%
      \centering
      \includegraphics[width=0.9\linewidth]{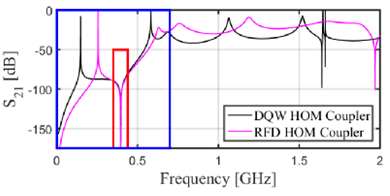}%
      \caption{}
      \label{fig:hl-lhc_crab_S21}%
       % }
   \end{subfigure} 
   \begin{subfigure}{0.45\textwidth}      
       % \subfloat[]{%
      \centering
      \includegraphics[width=0.9\linewidth]{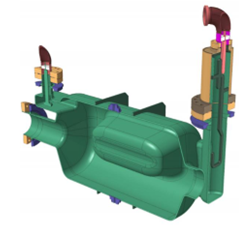}%
      \caption{}
      \label{fig:hl-lhc_crab}%
       % }\hfill
   \end{subfigure} \hfill
\caption{\(S_{21}\) (a) of the HOM couplers of DQW and RFD cavities for the HL-LHC project \cite{mitchell_hl-lhc_2022}. The latter is illustrated in (b).}
\end{figure}

When designing HOM couplers one has to take care to not extract power from the main mode. 
Figure~\ref{fig:hl-lhc_crab_S21} shows the transmission parameter \(S_{21}\) of the HOM couplers of two types of crab cavities for the HL-LHC project: 
double quarter-wave (DQW) and the radio frequency dipole cavities (RFD, see Fig.~\ref{fig:hl-lhc_crab}) \cite{mitchell_hl-lhc_2022}. 
One notices the notch filter characteristic at the frequency of the crabbing mode at 0.4~GHz, preventing this mode from propagating through the HOM couplers. 

\begin{figure}[h]
    \centering
    \begin{subfigure}{0.45\textwidth}
       \centering
       \includegraphics[width=0.9\linewidth]{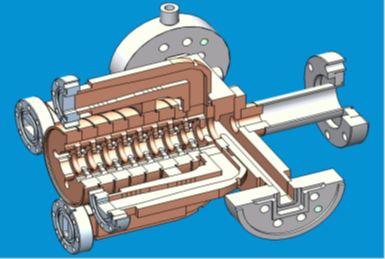}
       \caption{}
       \label{fig:couplers_dds}
    \end{subfigure}
    \begin{subfigure}{0.45\textwidth}
       \centering
       \includegraphics[width=0.9\linewidth]{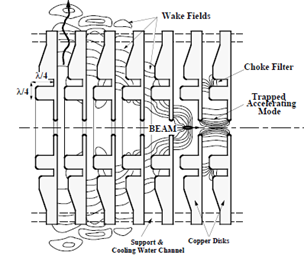}
       \caption{}
       \label{fig:couplers_choke}
    \end{subfigure}
    \caption{Examples of accelerating structures with HOM damping in each cell. (a) Part of a damped detuned structure for the Next Linear Collider \cite{jones_cas_2010}. (b) Part of a choke mode cavity for the C-band linear collider \cite{shintake_hom_1996}.}
\end{figure}

For long multicell accelerating structures it is not feasible to add couplers in the beam pipe only, as in the examples above. 
Often there are trapped modes which do not reach the ends of the structure. 
Therefore, one can add HOM couplers in some relevant cells, or even in each of them. 
%This is particularly advantageous for long detuned structures with trapped HOMs, i.e. modes which do not reach the beam pipes.
This is the case for the damped detuned structure designed for the Next Linear Collider (NLC) study. 
The extracted power is led through waveguides parallel to the structure axis to absorbers, as seen for a few cells in Fig.~\ref{fig:couplers_dds} \cite{jones_cas_2010}. 
In the case of the choke mode cavity conceived for the C-band e+e- linear collider study, the power is led out of each cell directly to absorbers as seen in Fig.~\ref{fig:couplers_choke}, achieving a stronger damping effect \cite{shintake_hom_1996}.

\begin{figure}[h]
    \centering
    \begin{subfigure}{0.38\textwidth}
       \centering
       \includegraphics[width=\textwidth]{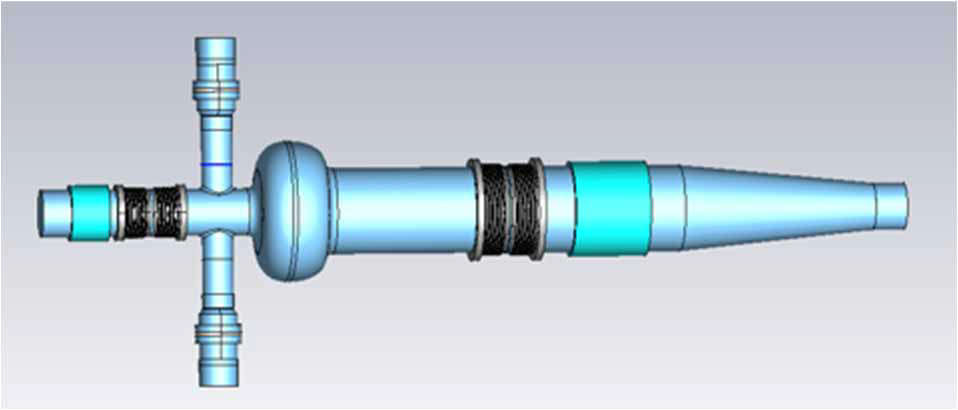}
       \caption{}
       \label{fig:bnl_esr}
    \end{subfigure}\hfill
    \begin{subfigure}{0.6\textwidth}
       \centering
       \includegraphics[width=\textwidth]{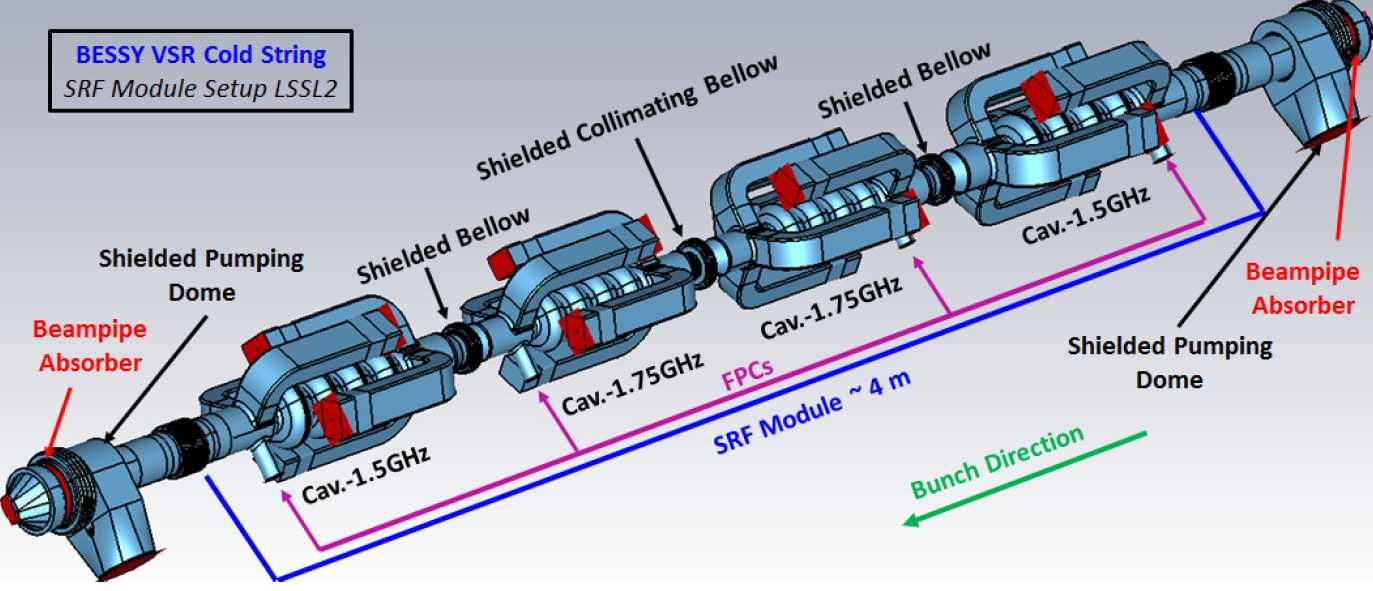}
       \caption{}
       \label{fig:bessy_vsr}
    \end{subfigure}
    \caption{Examples of beamline absorbers installed in a 1-cell cavity for the electron storage ring at BNL (a) \cite{rimmer_cavity_2022} and in the cavity string for BESSY's VSR (b) \cite{tsakanian:ipac18-wepml048}.}
\end{figure}

One drawback of HOM couplers is the relatively limited frequency of their damping range.
In many facilities, modes of higher frequencies  are also of concern, often coming into the single bunch effect range.
This is for example the case when the bunch length is very short, like in free electron lasers.
Therefore, in addition to couplers, many accelerator designs opt to add also broadband \textbf{absorbers} in the~beam pipes, for example made of ferrites or ceramics.
One example is the 1-cell cavity for the~electron storage ring at BNL shown in Fig.~\ref{fig:bnl_esr} \cite{rimmer_cavity_2022}.
One HOM beamline absorber is placed at either side.
Figure~\ref{fig:bessy_vsr} shows a cavity string designed for the BESSY Variable pulse-length Storage Ring (VSR)~\cite{tsakanian:ipac18-wepml048, glock:ipac18-thpmf033}. 
In this case  the~absorbers are placed before and after the whole string. 
The beam pipes are designed such that the HOMs of concern propagate through the whole structure and reach the absorber location.
The~accelerating mode is under the beampipe cutoff, so that it is not affected by the absorbers.

\subsection{Accelerator design}\label{accelerator_design}

Even with optimally designed cavities, HOMs are excited in accelerating structures, which may lead to an increase of the multi-bunch energy spread and transverse emittance.
Therefore, one has to carefully design the beam optics such that HOM effects are not dramatic.
This can be done for example by optimizing the recirculation arcs lattice in energy recovery linacs (ERLs) or by adding non-linear elements (see for example \cite{glukhov_multipass_2022}).

Each bunch in a multi-bunch beam will excite the same modes in a certain cavity. 
This is usually not dramatic, because in most cases the fields will not add up, but rather decohere.
It is not that unlikely though, especially in an accelerator with many accelerating structures, to have some modes resonantly excited by each bunch  in some cavities.
Also, there is a multitude of HOMs for multi-cell structures.
One straightforward thing to do is to check which modes are close to a beam resonance and are in danger to be excited resonantly, taking into account the fabrication errors.
At least one should avoid having the modes with larger loss factors close to a beam resonance.
One could \textbf{choose the bunch frequency} accordingly.
However this is usually not a parameter one can adjust, but one has to design the cavity such that such resonances are avoided, as explained in the previous sub-section.

In accelerators with many cavities, such as long linacs, there is a \textbf{spread in their HOM spectra} due to fabrication tolerances.
While the tolerances have to be kept small, some variation among the~structures is beneficial for the multi-bunch beam dynamics, because the effects on the bunches will not be cumulative.
With improved fabrication procedures there is a risk that this may no longer be the~case.

\begin{figure}[h]
   \centering
   \begin{subfigure}{0.45\textwidth}
      \centering
      \includegraphics[width=\linewidth,height=0.65\textwidth]{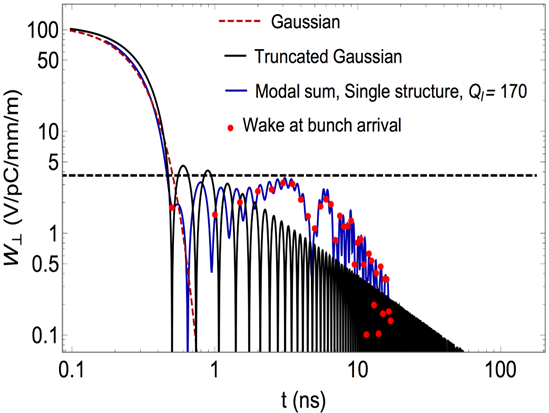}%
      \caption{}
      \label{fig:CLIC_DDS}
   \end{subfigure} \hfill
   \begin{subfigure}{0.45\textwidth}      
      \centering
      \includegraphics[width=\linewidth,height=0.65\textwidth]{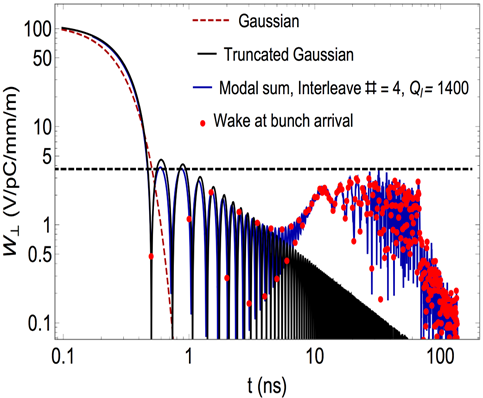}%
      \caption{}
      \label{fig:CLIC_DDS_interleave}%
   \end{subfigure}
\caption{Transverse wakefield calculated for a detuned structure for CLIC (a) and for the case of 4-fold interleaved structures (b) \cite{joshi_eucard2_2017}.}
\end{figure}

Another issue is that, due to the limited number of cells in the structure, the long-range wakefield recoheres after some time.
An example is given in Fig.~\ref{fig:CLIC_DDS} for a damped detuned structure (DDS) for the CLIC project study \cite{joshi_eucard2_2017, joshi:ipac19-weprb069}.
Due to detuning from cell to cell there is a frequency spread in each structure.
The HOM sum gives a wakefield function, which increases after a while, coming close to the wakefield limit given by beam dynamics calculations.
The loaded Q required to keep the wakefield below the beam dynamics limit is 170 in this case.
The effective number of cells can be increased by \textbf{interleaving}, i.e. making consecutive structures with slightly different frequencies.
Figure~\ref{fig:CLIC_DDS_interleave} shows an~example of the case of 4-fold interleaving, which relaxes the limit on Q to 1400.

\newpage

%========================================
%\input{sections/3_Compensate-HOMs}
\section{Methods to Compensate HOM Effects}\label{sec:compensate_homs}

Even with careful cavity and accelerator design, HOMs will still be excited and can affect the beam quality.
This section outlines strategies to mitigate their impact through fast feedbacks, beam alignment, and cavity retuning.

\subsection{Fast Feedbacks}\label{fast-feedbacks}
HOMs can induce variations in energy, arrival time, or transverse position from bunch to bunch which are unacceptable for the beam quality. Fast feedback systems measure these effects and correct them in real time.

\begin{figure}[h]
   \centering
   \begin{subfigure}{0.45\textwidth}
      \centering
      \includegraphics[width=0.8\linewidth]{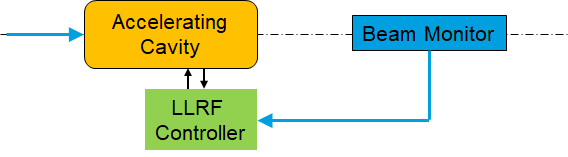}
      \caption{}
      \label{fig:long_feedback}%
   \end{subfigure} \hfill
   \begin{subfigure}{0.45\textwidth}      
      \centering
      \includegraphics[width=0.8\linewidth]{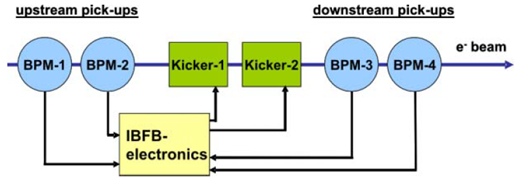}
      \caption{}
      \label{fig:transv_feedback}
   \end{subfigure}
\caption{Schematic diagram of a longitudinal (a) and a transverse (b) intra-bunch feedback system.}
\end{figure}

Figure~\ref{fig:long_feedback} shows the basic principle of a \textbf{longitudinal feedback} system. 
Dedicated beam monitors measure the arrival time (often also denoted as beam phase), energy, bunch length and other parameters of each bunch and send this information to a controller. 
This low level RF (LLRF) controller calculates corrections and feeds them back to the accelerating cavity (or set of cavities) so that the effect for the following beam pulse is corrected \cite{Pfeiffer:2019hna}.
This longitudinal feedback can be integrated into the beam loading compensation algorithm.
One option is to use a dedicated Bunch Arrival Corrector Cavity \cite{Pfeiffer:2019hna}.
%*** added citation

Transverse HOMs, usually dipole, sometimes also quadrupole and further modes, can cause a~large increase in the multi-bunch emittance.
Beam Position Monitors (BPMs) measure the resulting orbit deviations and send this information to a \textbf{intra bunch-train feedback} electronics \cite{schlott:epac06-thpch096}. 
This controls the~voltage of fast kickers which correct the offset of each bunch in the train (see Fig.~\ref{fig:transv_feedback}).

\subsection{Beam Alignment}\label{beam-alignment}

As mentioned in the previous chapter, transverse HOMs are not excited  by bunches aligned on the axis of the accelerating structure.
(Please note that the excited fields nevertheless often have transverse components along the axis and thus also affect subsequent, centred beams.)
Therefore one naturally comes to the conclusion that one way to avoid exciting dipole, quadrupole and other fields is to center the beam in the cavities.
The signal needed to do this is from the very modes that one wants to avoid.
Dipole modes are often the natural choice, due to their lower frequency and linear dependence on the beam offset.

\begin{figure}[h]
    \centering
    \includegraphics[width=0.5\linewidth]{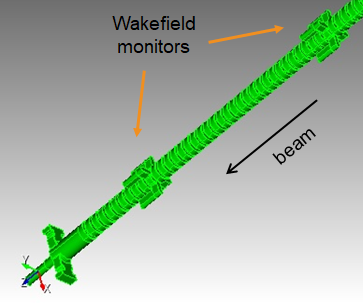}
    \caption{Detuned X-band structure with wakefield monitors \cite{dehler_x-band_ibic2013}.}
    \label{fig:X-band_structure}
\end{figure}

The principle is just like for a cavity BPM: the excited resonant fields are extracted by HOM couplers.
(Also the RF pickup designed for monitoring the accelerating fields may contain these modes.)
One or several dipole modes are filtered from the spectrum and monitored, preferably some with higher loss factor, since they are of main concern, but they also deliver higher signals.
Then one minimizes this signal by adjusting the beam offset with steerers.
This has been shown for TESLA cavities as mentioned in Ref.~\cite{baboi_homsc2016}.
In the case of the X-band structure shown in Fig.~\ref{fig:X-band_structure} \cite{dehler_x-band_ibic2013}, two so-called wakefield monitors have been specially designed to provide signals from the dipole modes inside the detuned structure, which allow one to infer the beam-to-structure alignment.

\subsection{Cavity Detuning and Retuning}\label{cavity-retuning}

Often there are just a few modes which alone may damage the beam quality significantly in case they are excited resonantly by a multi-bunch beam. 
Resonant amplification of HOMs occurs when their frequencies align with bunch harmonics. 

While one designs the accelerating cavities such that this should not happen, there is still a small probability that, due to fabrication spread, in some cavity one mode is shifted in frequency so that that it overlaps with a line in the beam spectrum.
In this case one has the option to detune the cavity and retune back the operating mode to the nominal frequency.
Because the cavity is inelastic, the HOMs will have slightly different frequencies than before.
The feasibility of this procedure was successfully tested experimentally.
In Ref.~\cite{yakovlev:ipac10-tupea020} the results after detuning a 1.3 GHz ILC-type cavity are given.
\newpage

%========================================
%\input{sections/4_Uses_ofHOMs}
\section{Uses of HOMs}\label{sec:hom_uses}

While HOMs are typically detrimental to accelerator performance, they can also serve valuable diagnostic purposes. 
This section explores a few ways in which HOM signals can be exploited to monitor cavity alignment and beam properties. 
The prerequisite is that one can monitor at least some of the HOMs, which is usually the case.

\subsection{Diagnostic of Beam Properties}\label{diagnostic-beam}

Dipole modes have a linear dependence on the transverse beam offset from the axis of the accelerating structure.
This makes it straightforward to use them as beam position monitors (BPMs), just like in cavity BPMs.
In this sense one just needs to install similar electronics as for dedicated beam monitors.
However, some peculiarities of the dipole modes in the accelerating cavities have to be taken into account.
For example, in multi-cell structures there is a multitude of modes of the same kind.
Separating one dipole mode may be difficult. 
One can also consider using several modes from the spectrum.
Also, keep in mind that when we talk about one dipole mode, there are actually two polarizations. 
For cavities with asymmetries, the two polarizations typically have different frequencies.
Disentangling them may be an~issue.
On the other hand, their polarization is often neither horizontal nor vertical.

\begin{figure}[h]
    \centering
    \includegraphics[width=0.9\linewidth]{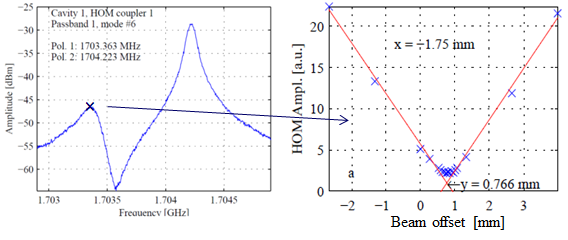}
    \caption{Dependence on the transverse bunch position (right plot) for one of the two polarizations of a dipole mode (left), measured at a TESLA cavity \cite{baboi_homsc2016}.}
    \label{fig:hom_bpm}
\end{figure}

Figure~\ref{fig:hom_bpm} shows on the left side the spectrum of a dipole mode with its two polarizations, measured at a TESLA cavity \cite{baboi_homsc2016}.
%*** added a couple of words
The right plot shows the dependence of the left peak when varying the transverse position of the bunch.
The linear dependence is clearly visible, enabling its use as a BPM.

%*** changed references to figures to `left' and `right'

Also other information on the beam can be extracted from HOM signals.
Monopole modes excited by the beam do not depend on the beam offset and come often as ``clean'' peaks in the HOM spectrum.
This straightforwardly leads to the idea of using them as beam phase monitors.
Usually, this parameter is measured with dedicated diagnostics, which is also used for feedback on the RF, as mentioned in the~previous section.
Reliable reference signals are crucial for this.
From HOM couplers one can get the~information on the beam phase from an excited higher order monopole mode together with the accelerating RF.
One can thus measure through the same cable directly the phase of the RF relative to the~beam, without an~additional reference signal.
While the measurement may not be as precise as a~dedicated monitor, it can be used to compensate for drifts in the machine.
This concept has been shown at TESLA cavities as described among others in Ref.~\cite{shi_beam_2018}.

\subsection{Diagnostic of Accelerating Structures}\label{diagnostic-structures}
Based on the linearity of dipole modes with the beam offset, one can also gain information on the structure alignment in the accelerator or even its straightness.

For superconducting cavities, which, once built inside a cryomodule, can no longer be accessed directly, the measurement of their transverse alignment after cooling is not trivial.
For this purpose, wire-based and optical systems have been developed for some projects \cite{rawnsley:linac04-mop89}.
An alternative is to use beam excited dipole modes.    
Based on the same principle as for determining the beam position, described in the previous sub-section, one can extract the cavity position with respect to a reference trajectory.
This has been measured for the cavities in several TESLA cryo-modules as described in Ref.~\cite{molloy_high_2007}.

\begin{figure}[h]
    \centering
    \includegraphics[width=0.5\linewidth]{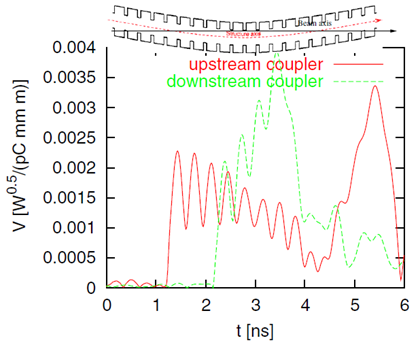}
    \caption{Simulated signal envelope from the upstream and downstream monitors of a X-band structure vs time assuming a bent structure \cite{dehler_x-band_2009}.}
    \label{fig:cell_alignment}
\end{figure}

For normal conducting, detuned, constant gradient structures one can also infer the transverse misalignment of the cells.
This relies on the fact that a certain mode in a dipole band is dominantly excited in one cell, whose resonant frequency corresponds to the synchronous one.
Therefore, the spectrum of a~dipole band carries the imprint of the cell misalignment.
A simulation example is shown in Fig.~\ref{fig:cell_alignment} for an X-band structure carrying so-called wakefield monitors \cite{dehler_x-band_2009} \cite{dehler_x-band_ibic2013}.

%========================================
%\input{sections/5_Summary}
\section{Summary}\label{sec:summary}

Higher Order Modes pose significant challenges to beam quality in accelerators. 
However, their effects can be effectively mitigated through careful design and operational strategies.
This lecture outlined various ways to address HOM-related issues, covering their avoidance and compensation.
Beneficial applications of HOM signals were also mentioned.

To avoid HOM excitation, the design of accelerating structures plays a critical role.
Optimizing cavity geometries to minimize loss factors for dangerous modes --- particularly those synchronous with the beam --- was discussed.
Additionally, HOM dampers, including couplers and broadband absorbers, are essential tools to reduce the quality factors of problematic modes without affecting the~accelerating mode.
Techniques like cell detuning and interleaving help to disrupt coherent mode buildup.

For compensating residual HOM effects during accelerator operation, fast feedback systems were presented as a robust solution.
Longitudinal intra-bunch feedbacks correct energy and timing variations, while transverse  feedbacks stabilize the beam orbit using kickers.
Beam alignment was also emphasized to suppress transverse mode excitation.
In cases where modes are resonantly excited in SC cavities, de- and retuning offers a remedy to shift HOM frequencies away from harmful harmonics.

Beyond mitigation, HOM signals can be used for diagnostics.
Dipole modes enable beam position and alignment monitoring, while monopole modes can help to achieve beam phase and energy stability.
Furthermore, HOM spectra can reveal structural misalignments, such as cell offsets in accelerating structures or cavity alignment within superconducting cryomodules.

In conclusion, addressing HOMs is important for a good beam quality.
Their mitigation requires a combination of proactive design and real-time correction. 
The beam-excited signals can be creatively exploited for diagnostics.

\printbibliography[title={\fontsize{14}{16}\selectfont References}] %Prints bibliography

\end{document}